\documentclass[12pt]{article}
\pdfoutput=1

\usepackage{amsthm,amsmath,natbib}
\usepackage{amssymb,amsfonts,bm,enumerate,graphicx}
\usepackage{xcolor,color}
\usepackage{booktabs}
\usepackage{caption}
\usepackage{subcaption}
\usepackage{float}
\usepackage{enumitem}
\usepackage{algorithmicx}
\usepackage{algorithm}
\usepackage{algpseudocode}
\usepackage{url}
\usepackage[margin=1in]{geometry} 
\usepackage[T1]{fontenc}
\usepackage[utf8]{inputenc}
\usepackage{fancyhdr}
\usepackage{setspace}

\usepackage[margin=1in]{geometry} 

\newcommand{\bs}{\bm{s}}
\def\bs{{\boldsymbol{s}}}

\def\iid{{\ {\buildrel \rm{iid}\over \sim}\ }}

\DeclareMathSymbol{\Real}{\mathbin}{AMSb}{"52}
\DeclareMathSymbol{\Natural}{\mathbin}{AMSb}{"4E}
\providecommand{\keywords}[1]{\textbf{\textbf{Keywords}} #1}

\makeatletter
\let\@fnsymbol\@arabic
\makeatother

\theoremstyle{plain}

\usepackage{natbib}
\title{Projecting Flood-Inducing Precipitation with a Bayesian Analogue Model}
\author{Gregory P. Bopp\thanks{Department of Statistics, Pennsylvania State University, 326 Thomas Building, University Park, PA 16802, U.S.A. Emails: gxb951@psu.edu, bshaby@psu.edu}, \, Benjamin A. Shaby\footnotemark[1], \,Chris E. Forest\thanks{Department of Meteorology and Atmospheric Science; Department of Geosciences; Earth and Environmental Systems Institute, Pennsylvania State University, University Park, PA 16802, U.S.A. Email: ceforest@psu.edu}, \,Alfonso Mej\'{i}a \thanks{Department of Civil and Environmental Engineering, Pennsylvania State University, University Park, PA 16802, U.S.A.}}
\date{}

\RequirePackage[colorlinks,citecolor=blue,urlcolor=blue]{hyperref}
\begin{document}
\maketitle
\begin{abstract}
The hazard of pluvial flooding is largely influenced by the spatial and temporal dependence characteristics of precipitation. When extreme precipitation possesses strong spatial dependence, the risk of flooding is amplified due to catchment factors that cause runoff accumulation such as topography. Temporal dependence can also increase flood risk as storm water drainage systems operating at capacity can be overwhelmed by heavy precipitation occurring over multiple days. While transformed Gaussian processes are common choices for modeling precipitation, their weak tail dependence may lead to underestimation of flood risk. Extreme value models such as the generalized Pareto processes for threshold exceedances and max-stable models are attractive alternatives, but are difficult to fit when the number of observation sites is large, and are of little use for modeling the bulk of the distribution, which may also be of interest to water management planners. While the atmospheric dynamics governing precipitation are complex and difficult to fully incorporate into a parsimonious statistical model, non-mechanistic analogue methods that approximate those dynamics have proven to be promising approaches to capturing the temporal dependence of precipitation. In this paper, we present a Bayesian analogue method that leverages large, synoptic-scale atmospheric patterns to make precipitation forecasts. Changing spatial dependence across varying intensities is modeled as a mixture of spatial Student-t processes that can accommodate both strong and weak tail dependence. The proposed model demonstrates improved performance at capturing the distribution of extreme precipitation over Community Atmosphere Model (CAM) 5.2 forecasts.
\end{abstract}


\section{Introduction} 
\label{sec:introduction}
In this paper, we develop a mixture model for daily precipitation allowing for varying spatial dependence for different storm magnitudes. Instead of constructing a parametric formulation of the atmospheric dynamics regulating precipitation, we develop a Bayesian analogue method that can accommodate complex temporal dependence patterns, wherein precipitation analogues are established by identifying similar synoptic-scale atmospheric patterns from a historical library of observed climate states that are associated with precipitation outcomes.

Global climate models (GCMs) are our main tool for forecasting future climate conditions, providing a means to assess potential changes in the frequency and magnitudes of events such as heat waves, droughts and floods, all of which can have profound impacts on human health. Unfortunately, the coarse spatial resolution of GCMs cannot resolve fine scale hydro-meteorological processes associated with precipitation extremes \citep{Boe06, Maraun10}. However, GCMs also produce smoothly varying atmospheric variables such as atmospheric pressure and temperature whose spatio-temporal patterns are closely linked with precipitation outcomes \citep{Xoplaki04, Raziei12}.

Analogue methods try to address the problem of making forecasts in the presence of complex temporal dependence without a model parameterization of any geophysical dynamics. In their simplest form, analogue methods match the current climate state to observed climate states from a library of historical observationsc in order to forecast some future quantity (e.g. precipitation tomorrow) with the observed successor of the historical match (e.g. precipitation on the day following the historical match). While the atmosphere is known to be a chaotic dynamical system that is unstable under slight perturbations of initial conditions \citep{Lorenz69}, analogue approaches are justified by the tendency of that system to regularly revisit subsets of the phase space over time. Analogue methods were originally developed as empirical tools for short-term weather forecasting \citep{Krick42} and climate modeling \citep{Barnett78}, but researchers have begun to recognize their utility in a variety of contexts, including machine learning \citep{Zhao16,Lguensat17}, wind-speed modeling \citep{Nagarajan15}, and air quality monitoring \citep{Delle14}. Historically, analogue methods have been empirical, somewhat ad-hoc tools, but recently there have been attempts at putting these into a probabilistic framework \citep{McDermott16,Mcdermott18}.

Precipitation has been the subject of extensive study as it plays a central role in agriculture, flood-risk, and hydrology. Distributional forecasts of precipitation are critical for water management, infrastructure design, and developing disaster preparedness strategies. While modern numerical forecasting weather models have achieved considerable success at making short term forecasts by approximating solutions to the complex atmospheric processes governing the generation of precipitation, they typically do not account for uncertainty in their inputs. The function of stochastic weather generators, in contrast, is not to make short-term forecasts, but to accurately reproduce the distributional characteristics of precipitation on a fine spatial grid, while quantifying the uncertainty associated with those estimates \citetext{see \citealp{Ailliot15} for a review}.

As precipitation is an inherently spatial phenomenon, several approaches have been proposed to model the spatial dependence features of both occurrences (presence/absence of rain) and intensities (positive rain accumulations). A popular stochastic weather generator was proposed in a seminal paper by \cite{Wilks98} that models precipitation across multiple sites in two parts: (1) a two-state Markov process controlling the occurrence of precipitation at a given site, and (2) a precipitation intensity model that accounts for spatial dependence between sties but ignores temporal dependence. \cite{Berrocal08} and \cite{Kleiber12} develop similar two-stage spatio-temporal models for observations from rain-gauge stations, wherein a latent Gaussian process is thresholded to model precipitation occurrence, while another marginally transformed Gaussian process controls precipitation intensity. While most models assume the same spatial dependence structure across precipitation intensities, \cite{Bardossy09} raise the issue of varying spatial dependence types across different intensities and aim to address it with an empirical copula approach. In this paper, we address the issue of varying spatial dependence types by modeling precipitation as a mixture of Student-t processes with different spatial correlation structures. Our approach to this problem is similar to that of \cite{Gelfand05}, who treat precipitation intensities as a Dirichlet process (DP) mixture of Gaussian processes. In the context of extreme value modeling, \cite{Fuentes13} have also explored DP mixtures of Gaussian processes to flexibly model spatial dependence, but in the context of modeling maxima, which support marginal transformations to generalized extreme value (GEV) margins. Gaussian processes and their mixtures, however, are characterized by weak tail dependence, and in the presence of strong spatial dependence among extremes, their application can lead to underestimation of flood risk. To allow for stronger tail dependence, similar approaches have been taken by \cite{Morris17} and \cite{Hazra18} who use skew-t processes and their mixtures, which possess strong tail dependence, to model ozone and fire risk extremes respectively. 

Since extreme precipitation stands to do the most damage, accurately capturing the spatial dependence for high intensities necessitates special consideration. The last decade has produced many new methods for modeling spatial extremes. Two common approaches to modeling spatial extremes are based on liming results: max-stable processes \citetext{see \citealp{Davison12} for a review}, which are appropriate for component-wise maxima over large temporal blocks, and generalized Pareto processes for high threshold exceedances \citep{Ferreira14}. However, full-likelihood inference for these models suffers from computational intractability when the number of spatial observation locations is large. While composite likelihood approaches have been proposed to remedy this issue \citep{Padoan10}, the scalability of full-likelihood methods remains an open problem. In addition to limiting models, several Bayesian hierarchical models that borrow ideas from classical geostatistics have been developed \cite{Cooley07,Sang09,Sang10, Reich12}. Unlike in the classical setup to modeling extremes, which requires a somewhat arbitrary qualification of an extreme event (e.g. the threshold in a peaks-over-threshold model or block size in a model for block maxima), our proposed mixture model incorporates the entire distribution of data, while accommodating different dependence types for different storm intensities.

In addition to accurately representing the spatial dependence, capturing temporal dependence has also been central to the field of precipitation modeling. Several Hidden Markov models for unobserved weather states have been proposed that aim to capture temporal dependence at various scales, including those attributable to large, synoptic-scale weather patterns \citep{Bellone00, Ailliot09, Flecher10}. Latent multivariate autoregressive models are also common approaches to modeling temporal dependence for both occurrence and intensity processes \citep{Bardossy92, Makhnin09, Rasmussen13}. Methods based in physics have also appeared in the statistical literature; recently, \cite{Liu18} proposed a Lagrangian advection reference frames approach to modeling storm dynamics that couples radar reflectivity and wind field data to make short-term precipitation forecasts.

In the remainder of the paper, we take up a similar aim to that of \cite{Gao14} and \cite{Gao17} who use analogue methods to couple GCM forecasts of predictive atmospheric variables with the historical precipitation so as to understand the changes in the distributions of extreme precipitation under different climate forcing scenarios. Our method is different from these earlier efforts in that it is a hierarchical Bayesian model based approach that makes use of the full data likelihood and can easily account for multiple sources of uncertainty. The following sections are outlined as follows: in Section \ref{sec:model_definitions}, we develop a Bayesian models for precipitation occurrences and intensities that capture temporal dependence with a probabilistic analogue formulation. Finally, in Section \ref{sec:precipitation_analysis}, we apply the model to precipitation data over the northeastern United States, and compare it with climate model and reanalysis distributional forecasts of extreme precipitation.

\section{Model Definitions} 
\label{sec:model_definitions}
One of the main challenges of modeling precipitation fields is that their distribution consists of a mixture of a preponderance of zeros and positive precipitation amounts. To account for this, the proposed spatio-temporal model for precipitation is made up of two parts: (1) an occurrence model for the presence versus absence of precipitation, and (2) an intensity model for positive precipitation amounts \citep{Wilks90,Berrocal08}. We begin by describing the occurrence model for precipitation.

\subsection{Occurrence Model} 
\label{sub:occurrence_model}
A common approach to modeling spatially varying, binary data is via data augmentation, wherein a continuous latent process is thresholded into two categories \citep{Albert93, Heagerty98, Collett02}. To model the point referenced, binary occurrence of precipitation, we use a Gaussian process for the unobserved, latent component, such that it is positive at a location $\bs$ whenever there is precipitation at $\bs$ and negative otherwise. This commonly used model is referred to as a clipped Gaussian process or spatial probit model \citep{DeOliveira00, DeOliveira18}. 

The spatio-temporal occurrence process $\{O_t(\bs), \,\bs \in \mathcal{S}\}, \, t =1, \ldots, T$, $\mathcal{S} \subset \mathbb{R}^2$, consists of spatial random fields that encode the presence versus absence of precipitation (1: presence, 0: absence) at a location $\bs$ and time $t$. It is defined in terms of a zero-thresholded, latent spatio-temporal Gaussian process $\{Z_t(\bs),\, \bs \in \mathcal{S} \}, \, t =1, \ldots, T$:
\begin{align*}
O_t(\bs) &= \begin{cases}
	1, &\text{ if } Z_t(\bs) > 0 \\
	0, &\text{ if } Z_t(\bs) \leq 0.
\end{cases}
\end{align*}
The latent processes $Z_t(\bs)$ (by abuse of notation) are parameterized by a mean function $\mu_t(\bs)$ and covariance function $C(\bs, \bs')$ that induce spatial dependence in the occurrence process to reflect the fact that nearby locations are more likely to have common presence or absence of rain than distant ones. Conditional on the mean and covariance functions, the spatially dependent processes are assumed to be independent in time, each distributed as:
\[Z_t(\bs)\overset{\mathrm{indep.}}{\sim} \mathrm{GP}(\mu_t(\bs), C(\bs, \bs')), \, \text{ for } t = 1, \ldots, T.\]
Temporal dependence is contained in the mean function. The covariance function can be expressed as a product of a variance parameter $\sigma^2_Z$ and correlation function $c(\bs, \bs')$ as $C(\bs, \bs') = \sigma_Z^2 c(\bs, \bs')$. However, since the variance term of this model is unidentifiable \citep{DeOliveira18}, it is fixed at $\sigma_Z^2 = 1$, making it sufficient focus on the spatial correlation function. Due to the flexibility provided by a parameter governing the smoothness of the process, we assume a correlation function from the general Mat\'{e}rn class \citep{Stein99}, which is both stationary and isotropic, making it expressible as a function of distance between locations $h = ||\bs - \bs'||$: $c_\nu(h) = \frac{2^{1-\nu}}{\Gamma(\nu)}\left(\sqrt{2\nu} \frac{h}{\rho}\right)^\nu K_\nu\left(\sqrt{2\nu} \frac{h}{\rho}\right),$ for range $\rho >0$ and smoothness $\nu >0$ parameters, where $K_\nu$ is a modified Bessel function of the second kind.

Using this construction, the marginal probability of occurrence at time $t$ and site $\bs$ can be expressed in terms of the mean function of the latent $Z_t(\bs)$ process as $\Pr(O_t(\bs) = 1) = \Pr(Z_t(\bs) > 0) = \Phi(\mu_t(\bs))$, where $\Phi$ is a standard normal distribution function. To allow for spatially varying marginal occurrence probability, the mean function is further modeled as a linear combination of $L$ spatial basis functions $\{\psi_l(\bs): \mathbb{R}^2 \rightarrow \mathbb{R}; \bs \in \mathcal{S}, \,  l= 1,\ldots,L \}$. For the spatial basis functions, we use Gaussian kernels although other choices are viable. For generic knot locations $\mathbf{v}_l \in \mathcal{S}, l = 1, \ldots, L$, the basis functions are defined as $\psi_l(\bs) = \phi\bigl(\frac{||\bs - \mathbf{v}_l||}{\Delta}\bigr)$, where $\phi$ denotes a standard Gaussian density function and $\Delta>0$ is a bandwidth parameter. Denoting the vector of basis functions at location $\bs$ as $\boldsymbol{\psi}(\bs) = (\psi_1(\bs), \ldots, \psi_L(\bs))'$, we model the mean function at time $t$ as a sum of an offset term $\gamma^{(O)}_t$ and a spatially varying term $\boldsymbol{\psi}(\bs)'\boldsymbol{\beta}^{(O)}_t$ as $\mu_t(\bs) = \gamma^{(O)}_t + \boldsymbol{\psi}(\bs)'\boldsymbol{\beta}^{(O)}_t$. For the vector of spatial basis coefficients $\boldsymbol{\beta}^{(O)}_t$, we assume an independent normal prior $\boldsymbol{\beta}^{(O)}_t \iid \mathrm{N}_L(\mathbf{0}, \sigma^2_{\beta^{(O)}}\mathrm{I}), \, t = 1, \ldots, T$. 

Since the presence or absence of precipitation is determined by whether $Z_t(\bs)$ is positive or negative, the $\gamma^{(O)}_t$ offset term can be thought of as governing the overall (non-spatially varying) probability of precipitation on day $t$. We will use this term to capture the temporal dependence in occurrence of precipitation by leveraging synoptic-scale climate forcings (e.g. geopotential height and temperature) that are concomitant with precipitation by construction of a prior on $\gamma^{(O)}_t$. We defer discussion of the analogue prior on $\gamma^{(O)}_t,\, t = 1,\ldots, T$ until Section \ref{sub:analogue_prior} as it is also used in the model for precipitation intensities. The priors for the Mat\'{e}rn dependence parameters are taken to be $\rho \sim \mathrm{Uniform}(\rho_l, \rho_u)$ and $\nu \sim \mathrm{Uniform}(0, 2)$, where in subsequent sections $\rho_l$ and $\rho_u$ are taken to be the minimum and maximum distance between observation locations.

\subsection{Intensity Model} 
\label{sub:intensity_model}
In this section, we develop a Bayesian model for the positive spatial precipitation intensity process. The intensity process is treated as a continuous process defined on the entire spatial domain, $\mathcal{S}$, that is masked by the occurrence process. In other words, the intensity process is only observed at locations where the occurrence process is positive. We make use of the Student-t process because of its flexibility in capturing heavy tailed behavior, which is commonly observed in precipitation data \citep{Vrac07, Naveau16}.

Denote the precipitation intensity process by $\{Y_t^*(\bs), \bs \in \mathcal{S}\}$ for times $t = 1, \ldots, T$. While precipitation intensities are strictly positive, it is much more convenient to work with a spatial process defined on the whole real line. The softplus function $f(x) = \log(\exp(x) -1)$, maps $f: (0, \infty) \rightarrow \mathbb{R}$ and is strictly increasing, preserving the natural ordering of the data. Both the softplus function and its inverse leave moderate to large values effectively unchanged. In the remainder of this section we define a model for the transformed precipitation intensities, $Y_t(\bs) \equiv f(Y_t^*(\bs))$. 

A location-zero Student-t process can be expressed as a Gaussian process scale mixture \citetext{see e.g., \citealp{Shah14}}: 
\begin{align*}
U(\bs) &= \sigma \epsilon(\bs) \\
\epsilon(\bs) &\sim \mathrm{GP}(\mathbf{0}, C(\bs, \bs')) \\
\sigma^2 &\sim \mathrm{IG}\left(\frac{a}{2}, \frac{a b}{2}\right)
\end{align*}
where $\mathrm{IG}(\frac{a}{2}, \frac{a b}{2})$ denotes a Inverse-Gamma distribution with shape $a/2$ and scale $ab/2$. After marginalizing over $\sigma$, $U(\bs)$ is a Student-t process with degrees of freedom $a$ and scale $b$. A Gaussian process is a limiting case of a Student-t process as $a \rightarrow \infty$.

To allow flexible spatial dependence types across different precipitation intensity levels, the transformed precipitation amounts $Y_t(\bs)$ are modeled as a finite mixture of Student-t processes. Finite mixtures can easily be accommodated via data augmentation. Let $\xi_t \in \{1, \ldots, K\}$ denote the latent mixture label for time $t$, where $K$ is the total number of mixture components. A multinomial logistic model, also referred to as the Luce model in the probabilistic-choice econometrics literature when modeling latent classes \citep{Luce59,Mcfadden73}, is used for the mixture class membership. Denote covariates (e.g., containing synoptic scale atmospheric information) by $\mathbf{u}_t \in \mathbb{R}^p$ for times $t = 1, \ldots, T$, vectors of coefficients for each class $k = 1, \ldots, K$ by $\boldsymbol{\alpha}_k \in \mathbb{R}^p$. Then, for linear predictor $\eta_{tk} = \mathbf{u}_t'\boldsymbol{\alpha}_k$, the probability $\pi_{t,k}$ that the process at time $t$ belongs to mixture component class $k$ is modeled as:
\begin{equation}
\pi_{tk} \equiv \Pr(\xi_t = k) = \frac{\exp\{\eta_{tk}\}}{\sum_{j = 1}^K \exp\{\eta_{tj}\}},  \, k = 1,\ldots, K; \, t = 1, \ldots, T.
\label{eq:mixture_prob}
\end{equation}
For identifiability, the loadings for the $K{\mathrm{th}}$ class are fixed $\boldsymbol{\alpha}_K = \mathbf{0}$ while the remaining loadings have independent normal priors: 
\[\boldsymbol{\alpha}_k \iid \mathrm{N}_p(\mathbf{0}, \sigma^2_\alpha\mathrm{I}_p),\quad k = 1, \ldots, K - 1.\]
Instead of using a common spatial dependence structure, each mixture class is free to have different spatial covariance, degrees of freedom $a_k$, and scale $b_k$ parameters. Any spatial variation in the location surface at time $t$ is captured by an offset term and a linear expansion of spatial basis functions $\mu_t(\bs) = \gamma^{(I)}_t + \boldsymbol{\psi}(\bs)'\boldsymbol{\beta}^{(I)}_t$, just as was done in Section \ref{sub:occurrence_model}. Conditional on the mixture label at time $t$, $\xi_t = k$, the intensity process is modeled as follows: 
\begin{align*}
Y_t(\bs) &= \gamma^{(I)}_t + \boldsymbol{\psi}(\bs)'\boldsymbol{\beta}^{(I)}_t + \sigma_t \epsilon_t(\bs) \\
\epsilon_t(\bs) &\sim \mathrm{GP}(\mathbf{0}, c_k(\cdot, \cdot))\\
\sigma_t^2 &\overset{indep}{\sim} \mathrm{IG}\left(\frac{a_k}{2}, \frac{a_k b_k}{2}\right).
\end{align*}

For each mixture class, we assume an isotropic Mat\'{e}rn correlation function with potentially different smoothness $\nu_k$ and range $\rho_k$ parameters:
\[c_k(h) = \frac{2^{1-\nu_k}}{\Gamma(\nu_k)}\left(\sqrt{2\nu_k} \frac{h}{\rho_k}\right)^{\nu_k} K_{\nu_k}\left(\sqrt{2\nu_k} \frac{h}{\rho_k}\right), \, k = 1, \ldots, K. \] 
Just as in Section \ref{sub:occurrence_model}, we assume independent normal priors for the spatial basis coefficients $\boldsymbol{\beta}^{(I)}_t \iid N(\mathbf{0}, \sigma^2_\beta\mathrm{I}), \, t = 1, \ldots, T$, and we use an analogue prior for $\gamma^{(I)}_t, \, t = 1, \ldots, T$, which we discuss in Section \ref{sub:analogue_prior}. The following priors are used for the degrees of freedom and scale parameters: $a_k \iid \mathrm{Uniform}(0, 30), \, k = 1, \ldots, K$, and $b_k \iid \mathrm{Gamma}(0.1, 10)$, where the Gamma distribution is parameterized with shape and scale respectively. Priors for the Mat\'{e}rn covariance parameters $\rho_k$ and $\nu_k$ are assumed to be indpendent across $k$ and the same as those in the occurrence model.

Because we use a mixture of Student-t processes, different mixture components are capable of capturing different spatial dependence types, allowing for varying dependence strengths across different intensities. The upper tail dependence of a generic stationary and isotropic spatial process $\{W(\bs), \bs \in \mathcal{S}\}$ can be characterized by the tail dependence function at level $u \in (0,1)$, defined as:
\begin{equation}
\chi_u(h) = \Pr\biggl\{W(\bs + h) > F^{-1}(u)|W(\bs) > F^{-1}(u)\biggr\},
\label{eq:chiu}
\end{equation}
where $F$ denotes the marginal distribution function of $W(\bs)$. The limit $\chi(h) = \underset{u \rightarrow 1}{\lim}\chi_u(h)$ determines the tail dependence class. The process is said to be asymptotically independent at distance $h$ is $\chi(h) = 0$, and asymptotically dependent at distance $h$ otherwise. Gaussian processes are asymptotically independent processes with $\chi(h) = 0$ for all $h > 0$, making them suitable for modeling physical phenomena exhibiting weak tail dependence \citep{Sibuya60}. However, Student-t processes and their finite mixtures possess asymptotic dependence making them suitable in the case of strong tail dependence, but less so when the data exhibit weakening dependence at higher levels \citep{Nikoloulopoulos09}. Because a Gaussian process is a limiting case of the Student-t process, a Student-t process with large degrees of freedom is capable of capturing relatively weak dependence at sub-asymptotic levels. 

\subsection{Analogue Prior}
\label{sub:analogue_prior}
Both models described in Sections \ref{sub:occurrence_model} and \ref{sub:intensity_model} possess location offset terms $\gamma_t^{(O)}$ and $\gamma_t^{(I)}$, which here we will denote generically as $\gamma_t, \, t = 1, \ldots, T$. The priors for the offset parameters have thus far been left unspecified. In this section, we describe how the similarities between climate forcings at different times can be used to model the temporal dependence in the offset term. Following \cite{McDermott16} and \cite{Mcdermott18}, who take a similar approach to forecasting soil moisture and waterfowl settling behavior, we will refer to this as an analogue prior on $\gamma_t, \, t = 1, \ldots, T$.

The presence and intensity of precipitation is closely related to other atmospheric conditions such as atmospheric pressure, temperature, and water vapor. Rather than explicitly model the complex precipitation dynamics, we use closely related atmospheric variables to identify historical analogues (times $t', \, t' \neq t$) of the atmospheric conditions at time $t$. Because of concomitance of atmospheric conditions with precipitation, the precipitation conditions during identified historical analogue times can then be used to inform the precipitation for the reference time. In a classical analogue model \citep{Barnett78,Sugihara90}, identified historical analogue precipitation fields and their weighted averages would be used as the forecasts for the reference time. Instead, we make a weaker assumption, and only borrow information about the location offset terms $\gamma_t$ from analogue precipitation fields.

\sloppy To formalize this, for each time $t = 1, \ldots, T$, define a vector of weights 
$\mathbf{w}_t = \allowbreak(w_{t,1}, \ldots, w_{t,t-1}, w_{t, t + 1}, \ldots, w_{t, T})$, quantifying the similarity between the atmospheric conditions at time $t' \neq t$ and those at time $t$. For identifiability, the elements of the weight vectors are restricted to $w_{t, t'} \in [0,1]$, $w_{t,t} = 0$, and $\sum_{t' \neq t} w_{t,t'} = 1$,  for all $t$. Combining the location parameters for other times $t'\neq t$ into a single vector $\boldsymbol{\gamma}_{-t} = (\gamma_1, \ldots \gamma_{t-1}, \gamma_{t+1}, \ldots, \gamma_T)'$, we specify conditional normal priors on the location parameters $\gamma_t| \boldsymbol{\gamma}_{-t} \sim N(\boldsymbol{\gamma}'_{-t}\mathbf{w}_t, \sigma_\gamma^2), \, t = 1, \ldots, T$, so that the prior mean is a weighted average of the precipitation location parameters from historical conditions with strong similarity to the reference conditions.

The weights are calculated using a kernel function applied to distances atmospheric conditions at different times, such that more similar historical atmospheric conditions receive larger weights. , we use an unnormalized, compact Gaussian kernel function: $g(d; \theta, \tau) = \exp\{\frac{-d^2}{2 \theta}\}\mathrm{1\{d < \tau\}}, \, d > 0$, where $d$ is a measure of distance, $\theta$ is a kernel bandwidth parameter, and $\tau$ is a threshold ensuring that the kernel is compact, which prevents irrelevant non-analogue days from receiving positive weight. In subsequent sections it is fixed such that on average (across times $t$) only the top $m$ nearest analogues receive weight. When the kernel bandwidth is large, many historical analogues will contribute a small weight, and when the bandwidth is small, fewer historical analogues will contribute a larger weight. Given distances $d_{t,t'}$ between all times $t', \, t' \neq t$ and $t$, unnormalized weights $w^*_{t,t'}$ are calculated as $w^*_{t,t'} = g(d_{t,t'};\theta, \tau)$, and normalized to give weights $w_{t,t'} = w^*_{t,t'}/\sum_{j\neq t} w^*_{t,j}$.

Due to the high-dimension of geopotential height fields and temperature fields, some dimension reduction is needed before calculating distances between atmospheric covariates. To reduce the dimension while preserving the relevant variation in the data, each of the atmospheric variables are projected onto the first several components of the empirical orthogonal functions (EOFs) calculated from the data \citep{Jolliffe02,Hannachi07,Demsar13}. Rather than rely on snapshots of the atmospheric conditions on any given day, we employ the time lagging approach of \cite{Takens81}. Combining lagged EOF loadings into trajectories of atmospheric conditions over time gives a fuller reconstruction of the state-space of the dynamical system \citep{Sugihara90}. Let $\mathbf{x}_t$ denote the loadings for the first several components of relevant EOFs on day $t$, we construct a new embedding matrix as $X_t = [\mathbf{x}_t, \mathbf{x}_{t-1}, \ldots, \mathbf{x}_{t-r}]$, where $r$ is the number of lagged time steps. Doing this for all time points, the quality of $Y_{t'}$ as an analogue for $Y_t$ is determined by the distance (e.g., Euclidean norm) $d_{t,t'} = ||X_t - X_t'||$ between $X_t'$ and $X_t$. 

For both models, posterior samples are made using Markov chain Monte Carlo (MCMC), the details of which can be found in Appendix \ref{sec:mcmc_details}. Gibbs updates are available for some parameters, and the remaining are made with Metropolis-Hastings updates. Samples of the parameters are made for all observation times, and in-sample posterior predictive draws can be made directly, e.g. $Y_t(\bs')|\mathbf{Y}_{1:T}$ for $t \in \{1,\ldots, T\}$ and non-observation location $\bs' \notin \{\bs_1,\ldots, \bs_n\}$. However, to make out-of-sample posterior predictive draws, for example of $Y_{t'}|\mathbf{Y}_{1:T}$ for time $t' > T$, distances between atmospheric conditions on the future day and historical days must first be calculated, e.g. $d_{t',t}$ as well as their corresponding normalized weights $w_{t',t}$ from $w^*_{t',t} = g(d_{t',t};\theta, \tau)$ for $t = 1, \ldots, T$, conditional on posterior samples of $\theta$. 

To assess the utility of the model two simulation studies are performed, the results of which are summarized in the Supplementary Material.

\section{Precipitation Analysis} 
\label{sec:precipitation_analysis}
\begin{figure}[t!]
  \begin{center}
		\includegraphics[width=1.1\textwidth]{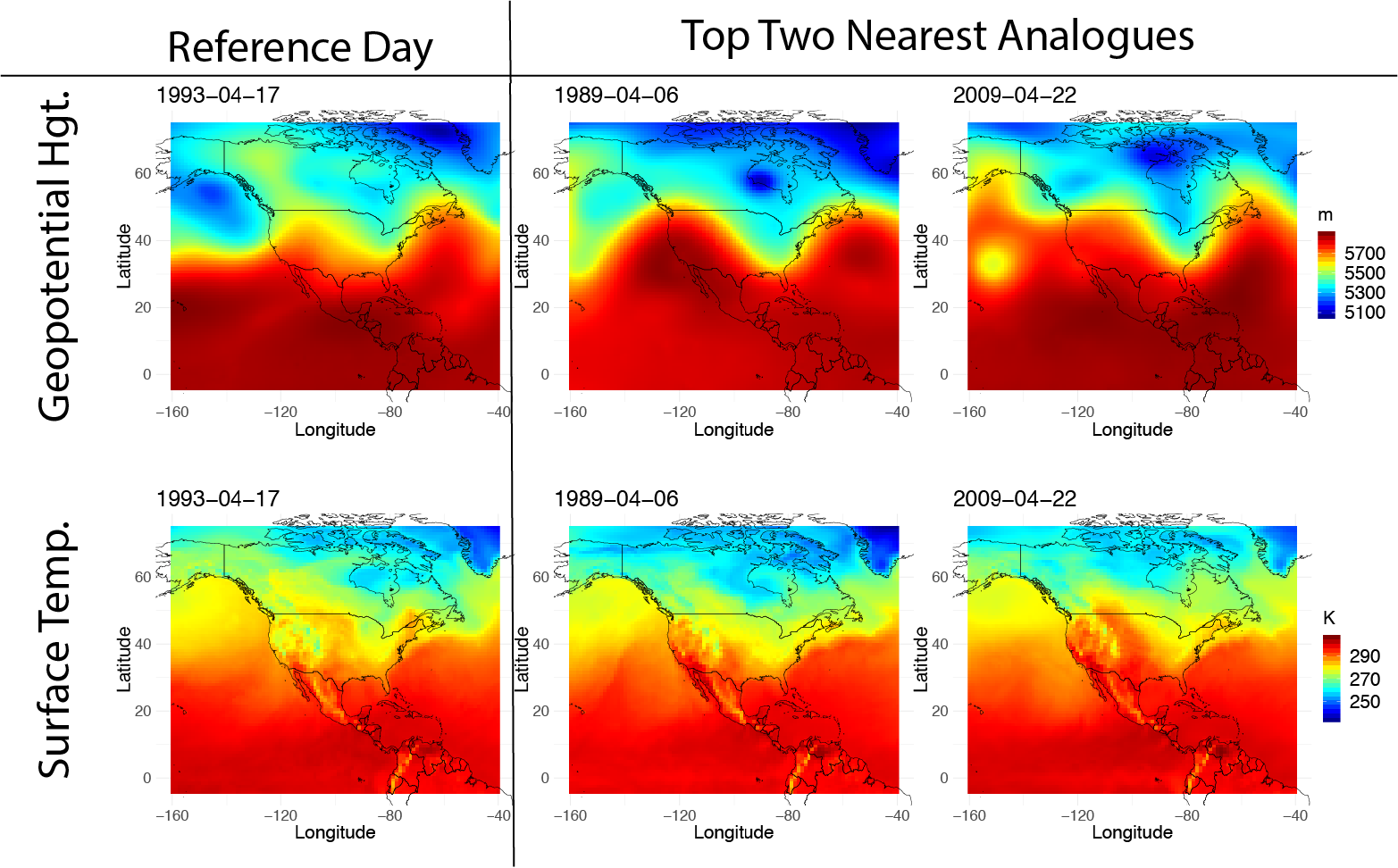}
		\caption{The 500 hPa geopotential heights and surface temperatures for a reference day are show in the leftmost column. The middle and right columns show the corresponding fields for the top two most similar days based on the Euclidean distance between lagged vectors of the top 10 PC loadings of both variables.}
		\label{fig:analogue_fields}
	\end{center}
\end{figure}

In this section, we apply our model to daily precipitation observed over the Susquehanna river basin in southern New York and Pennsylvania. By coupling daily precipitation accumulations observed at rain gauge stations with predictive atmospheric conditions, we can apply the proposed analogue model to make forecasts of precipitation over watersheds. The rain gauge data come from the National Oceanic and Atmospheric Administration (NOAA) (\href{https://www.ncdc.noaa.gov/ghcnd-data-access}{https://www.ncdc.noaa.gov/ghcnd-data-access}) and consist of daily precipitation accumulations (in mm) observed between 1986 and 2017 for $n = 174$ gauge stations. To capture the temporal dependence, meteorological reanalysis estimates of geopotential height and surface temperature are used to construct the distance matrix for the analogue prior on location terms in the occurrence and intensity models. Reanalysis computer models infill meteorological fields on a spatial grid by assimilating historical, spatially varying atmospheric observations, which are treated as boundary conditions in a consistent model of the climate system. We consider 500 hPa geopotential height and surface temperature from the Modern-Era Retrospective analysis for Research and Applications, Version 2 (MERRA-2) project \citep{Gelaro17} for identifying analogues. The analogue prior distance matrix is constructed using the lagged EOF loading approach described in Section \ref{sub:analogue_prior} with $L = 3$ lagged time steps and the first 10 PCs of each of the two MERRA-2 atmospheric variables. An example of the 500 hPA and surface temperature fields for reference and nearest analogue days are shown in Figure \ref{fig:analogue_fields}. 

We evaluate three models for daily precipitation: occurrence and intensity models with (M1) independence priors on $\gamma^{(O)}_t$ and $\gamma^{(I)}$, (M2) analogue priors on $\gamma^{(O)}_t$ and $\gamma^{(I)}$ with distance matrix calculated using MERRA-2 atmospheric variables over the continental United States (CONUS), and (M3) analogue priors on $\gamma^{(O)}_t$ and $\gamma^{(I)}$ with distance matrix calculated using MERRA-2 atmospheric variables over the a region surrounding Pennsylvania (local PA). The data is split into training (1986-2000) and holdout periods (2001-2017), during each year of which we consider only the months April, May, and June. During the training period, 20 stations are also held out for model evaluation of in-sample prediction (interpolation). Inference on each is performed with a Metropolis within Gibbs Markov chain Monte Carlo (MCMC) algorithm run for $100,000$ iterations. The models are evaluated by two measures: by their ability to capture the distribution of precipitation when making (1) out-of-sample forecasts and (2) in-sample, kriging predictions. To evaluate (1), out-of-sample posterior predictive draws are made from each model during holdout period at each of the 174 gauge locations, and to evaluate (2), in-sample posterior predictive draws are made at holdout stations for each day of the training period. In addition to these three models, we also consider the distributional forecasts of precipitation from the LENS historical run during the holdout period and MERRA-2 estimates during the training period. LENS fields of 500 hPa geopotential height and surface temperature are projected onto the MERRA-2 EOF basis for comparability of fields when calculating distances. Since reanalysis estimates effectively condition on observed the atmospheric boundary conditions to infill precipitation, we also compare the in-sample, kriging predictions of the proposed model to MERRA-2 reanalysis estimates of precipitation. Both MERRA-2 and LENS precipitation forecasts are made on a grid. To resolve the spatial mismatch between grid cells and gauge locations, gauge station forecasts for the LENS and MERRA-2 models are made by assigning each gauge station the precipitation amount from its nearest LENS and MERRA-2 grid cell.

The models are evaluated based on how well they capture the distributional characteristics of extreme precipitation. For each day of the holdout period, the maximum daily precipitation accumulation across all stations is calculated. The distribution of maxima is then compared to the predictions from each model using tail weighted continuously ranked probability scores (TWCRPS) \citep{Gneiting14}. For a single sample $y$, the TWCRPS is calculated as:
\begin{equation}
\mathrm{TWCRPS}\{\hat{F}, y\} = \int_{-\infty}^{\infty}(\hat{F}(x) - 1\{y \leq x\})^2 w(x) dx \, ,
\label{eq:twcrps}
\end{equation}
where $\hat{F}$ refers to the model estimate of the target distribution from which $y$ is drawn, and $w(x)$ is a weight function, which for this analysis we take to be $w(x) = 1\{x \geq q_{0.8}\}$ where $q_{0.8}$ is the $0.8$ quantile of the simulated total precipitation data. In practice, as we do here, the TWCRPS is averaged over a sample $y_1, \ldots, y_{\tilde{T}}$ as $\mathrm{TWCRPS} = \sum_{i=1}^{\tilde{T}} \mathrm{TWCRPS}\{\hat{F}, y_i\}$ (where $\tilde{T} = T$ for in-sample prediction and $\tilde{T}$ is the number of holdout days for out-of-sample prediction). Lower TWCRPS scores correspond to better correspondence between the empirical and model-based distributions. We consider two TWCRPS weight functions: $w_1(x) = 1\{x \geq q_{0.5}\}$ and $w_2(x) = \Phi\{(x-q_{0.5})/s_{emp}\}$ where $q_{0.5}$ and $s_{emp}$ are the sample median and standard deviation of the observed precipitation. The TWCRPS results are summarized in Table \ref{tab:twcrps_rcp85_oos}. Both versions of the analogue prior model outperform the independence prior model in capturing the distribution of precipitation extremes during the holdout period. Interestingly, the model with analogue distance matrix constructed from locally defined EOFs around PA shows greater predictive skill than both the one constructed from CONUS EOFs and the LENS model forecasts. Moreover, the in-sample forecasts of precipitation using the Local PA analogue prior model are also competitive with the MERRA-2 model. As such, we focus the remainder of our analysis on the analogue prior model using local PA fields.

\begin{figure}
    \centering
    \begin{subfigure}[b]{0.48\textwidth}
        \includegraphics[width=\textwidth]{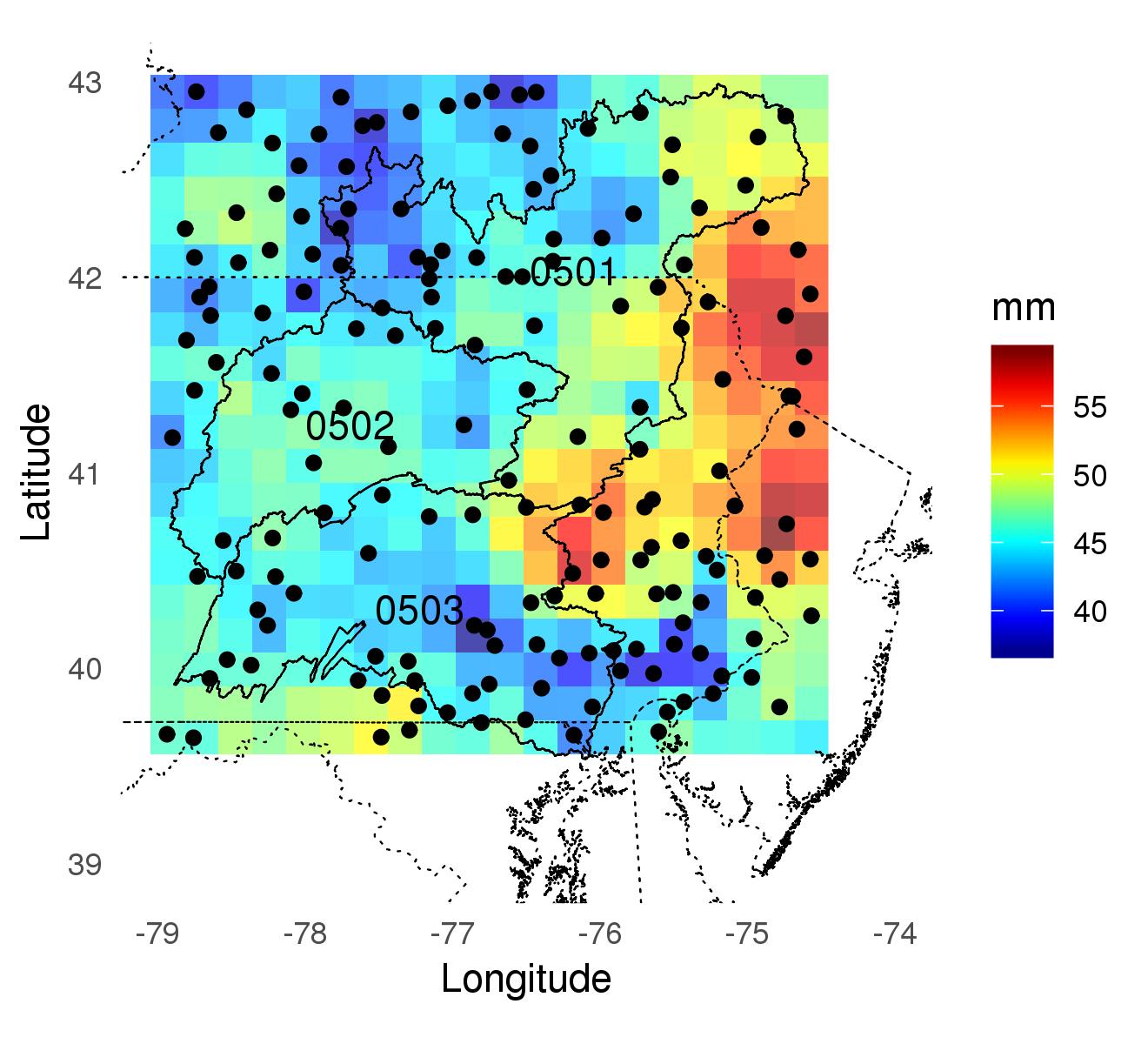}
    \end{subfigure}
    ~ 
    \begin{subfigure}[b]{0.48\textwidth}
        \includegraphics[width=\textwidth]{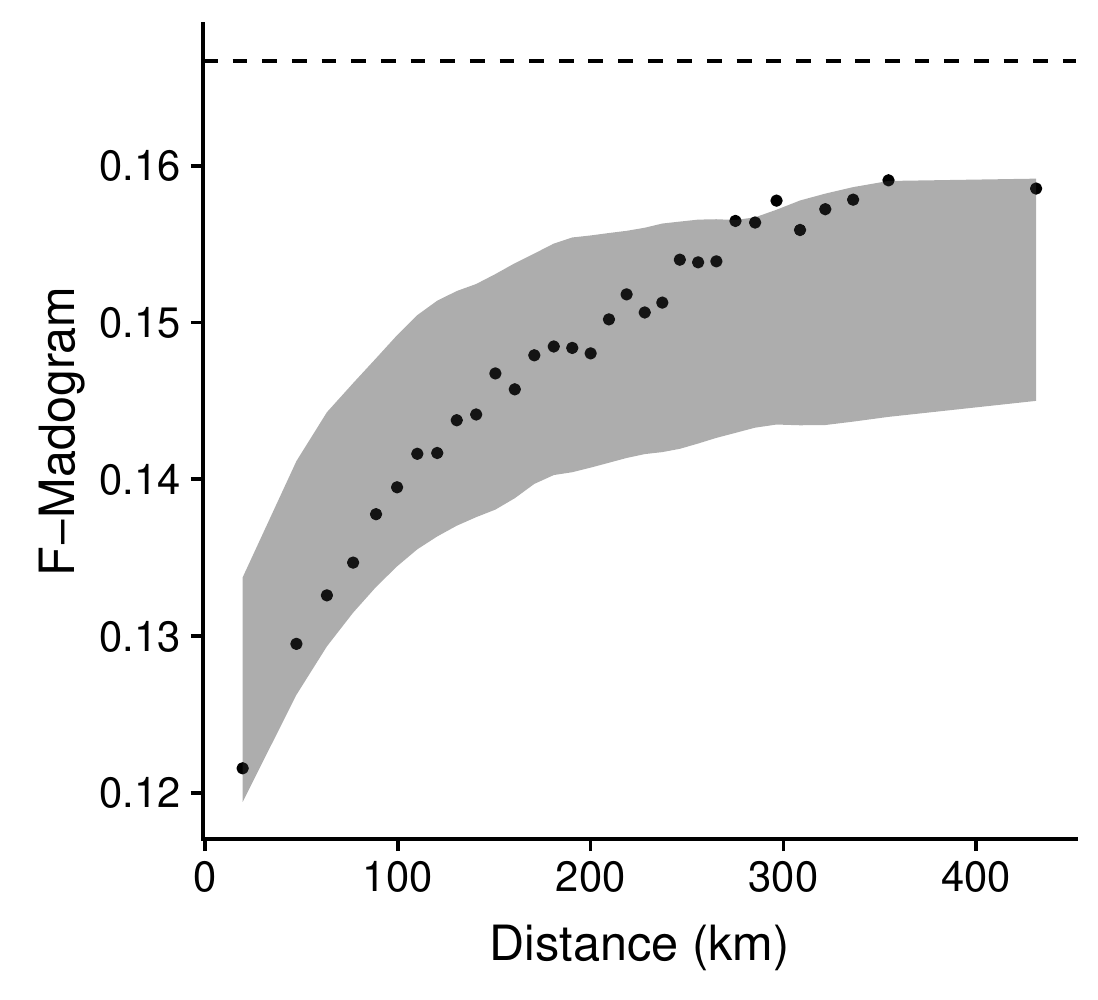}
    \end{subfigure}
    \caption{Left: Pointwise estimates of the $99^{\mathrm{th}}$ percentile of daily precipitation over the observation region based on the fitted analogue prior model overlaid with upper (0501), western (0502), and lower (0503) branch Susquehanna drainage basin boundaries. Right: Empirical F-madogram estimates (points) and 95\% credible intervals (ribbons) for monthly maxima of daily precipitation accumulations. A dashed horizontal line corresponding to independence is plotted for reference.}	
    \label{fig:q990_marginal}
\end{figure}

\begin{table}[t!]
\begin{center}
\caption{Tail weighted continuously ranked probability scores (TWCRPS) for the Student-t process mixture, LENS model forecasts of the distribution of daily maximum precipitation among all rain gauge locations during the holdout period (2001-2017), and MERRA-2 model forecasts of the distribution of daily maximum precipitation among holdout gauge locations during the training period (1986-2000). The better value for each criteria is given in bold. Estimated standard errors are given in parentheses.}
\label{tab:twcrps_rcp85_oos}
\resizebox{\linewidth}{!}{
\begin{tabular}{cc|ccccc}
\toprule
\multicolumn{1}{c}{\bfseries Predictions}&\multicolumn{1}{c}{\bfseries TW Fun.}&\multicolumn{1}{c}{Indep. Prior}&\multicolumn{1}{c}{ CONUS}& \multicolumn{1}{c}{Local PA} &\multicolumn{1}{c}{LENS} & \multicolumn{1}{c}{MERRA-2}\tabularnewline
\hline
In-sample & $w_1(x)$& 7.43  (7.35, 7.55)& 7.41 (7.32, 7.54)& 7.37 (7.28, 7.48) & & \textbf{7.30} (-,-)\\
In-sample & $w_2(x)$& 6.66 (6.59, 6.77)& 6.64 (6.56, 6.77)& 6.61 (6.53, 6.70) & & \textbf{6.54} (-,-)\\
Out-of-sample & $w_1(x)$ &20.9 (13.1, 25.2)&16.3 (15.2, 17.4)&\textbf{11.8 (11.2, 12.5)}&13.7 (-,-)&  \\
Out-of-sample & $w_2(x)$ &19.6 (12.0, 23.7)&15.0 (14.0, 16.1)&\textbf{10.7 (10.2, 11.4)}&12.7 (-,-)& \\
\bottomrule
\end{tabular}
}
\end{center}
\end{table}
A map of the pointwise, marginal $99^{\mathrm{th}}$ percentile estimates of the posterior predictive distribution of daily precipitation over the study region are given in Figure \ref{fig:q990_marginal}. Estimated precipitation levels tend to be highest over the northeastern part of the state as well as over parts of New Jersey and New York. To assess the correspondence between the empirical and model estimates of the spatial pattern of tail dependence, we also examine the F-madogram for monthly maxima \cite{Cooley06}. The F-madogram is analogous to the more traditional madogram from classical geostatistics, but is guaranteed to exist even when the first moment of the process under consideration is undefined. For a generic, stationary and isotropic spatial process $\{W(\bs), \bs \in \mathcal{S}\}$, $\mathcal{S} \subset \mathbb{R}^2$, with marginal distribution function $F$, and spatial lag $h > 0$, the F-madogram, defined as
\begin{equation}
\lambda_F(h) = \frac{1}{2}\mathrm{E}\biggm| F\{W(\bs')\} - F\{W(\bs)\}\biggm|, \, \text{ for } ||\bs - \bs'|| = h\,,
\label{eq:fmadogram}
\end{equation}
and describes the dependence in the pairs $(W(\bs), W(\bs'))$ that are a distance $h$ apart. When the pair are perfectly dependent, $\lambda_F(h) = 0$, and when the pair are independent, $\lambda_F(h) = 1/6$. The figure shows both good correspondence between the empirical and model estimates of $\lambda_F(h)$ as well as apparent non-zero dependence of monthly maxima even at spatial lags of $h = 400$ km, which suggests that the model is capturing the spatial dependence properties of extreme precipitation well.

\begin{figure}[t!]
  \begin{center}
		\includegraphics[width=\textwidth]{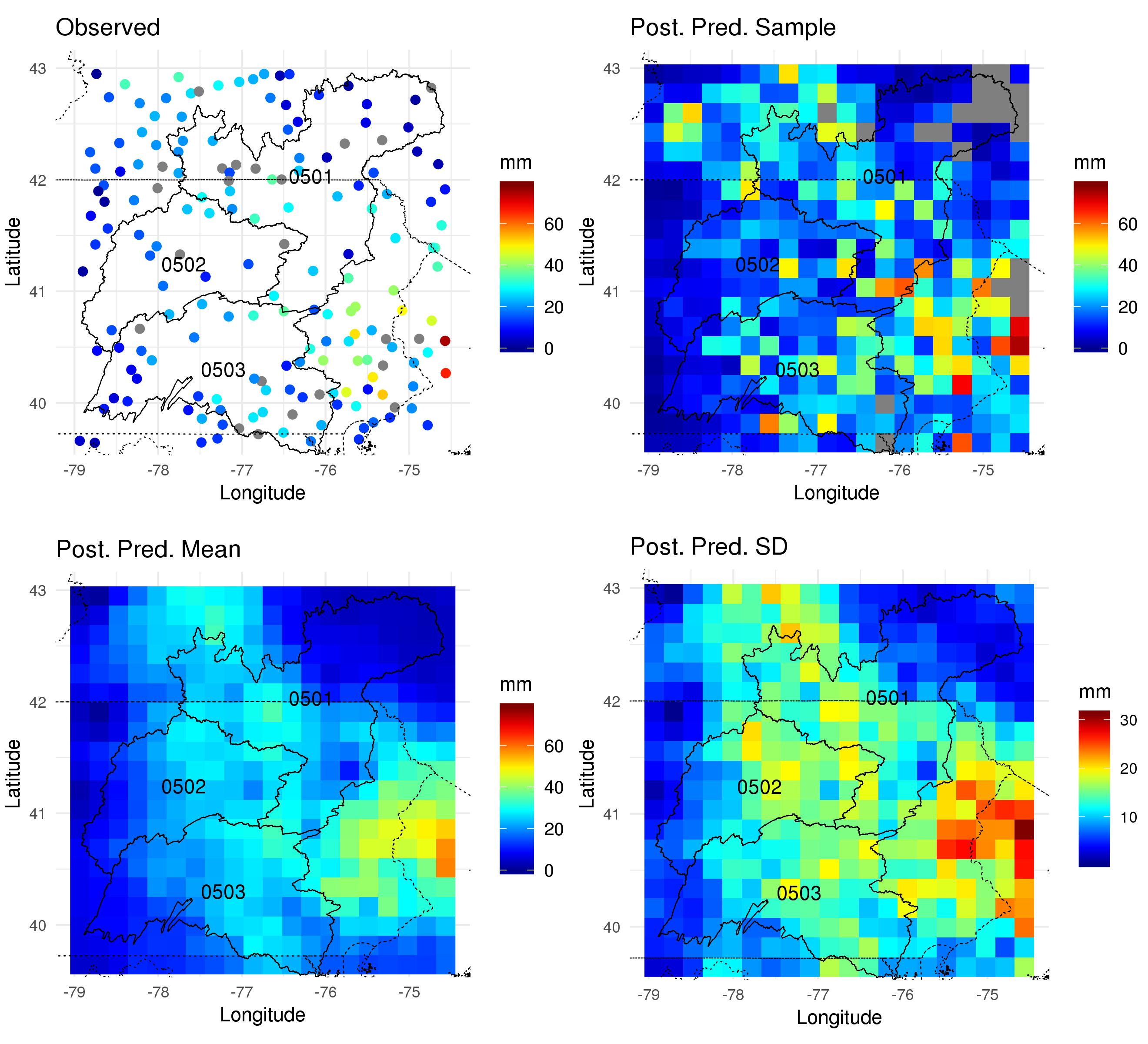}
		\caption{Observed daily precipitation amounts at gauge locations (top left), posterior predictive draw (top right), posterior predictive mean (bottom left), and posterior predictive standard deviation (bottom right) for a single day during the observation period. Grey points and cells correspond to zero precipitation amounts. The general spatial mean surface as well as degree of smoothness in the observed precipitation data is well captured by that of the posterior predictive sample.}
		\label{fig:precip_ppred}
	\end{center}
\end{figure}

\begin{figure}[t!]
  \begin{center}
		\includegraphics[width=1.1\textwidth]{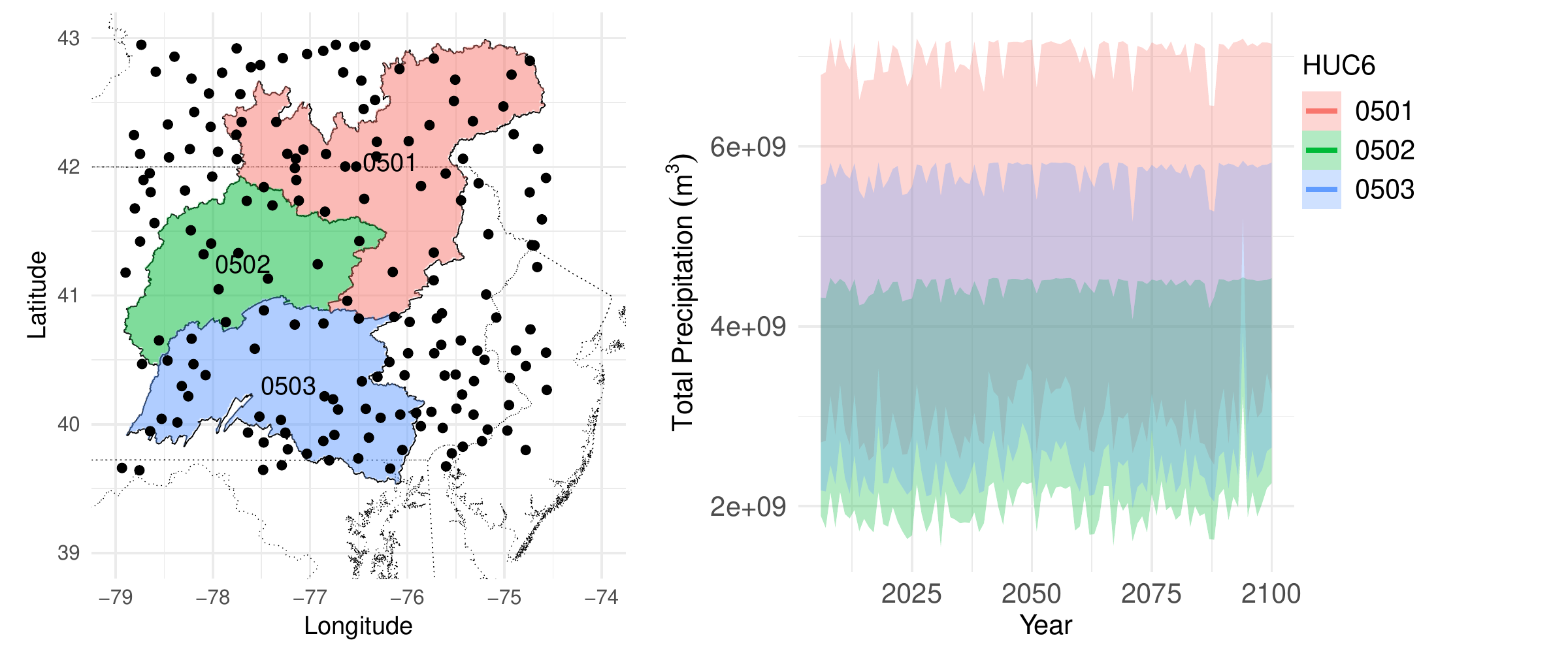}
		\caption{Global Historical Climate Network (GHCN) daily gauge station locations over Pennsylvania and surrounding states are overlaid with upper (0501), western (0502), and lower (0503) branch Susquehanna drainage basin boundaries (left). Forecasted yearly (during April-May-June) maximum total precipitation 95\% credible intervals for each basin are plotted on the right.}
		\label{fig:rcp85_huc6_forecast}
	\end{center}
\end{figure}

For illustration of the model based forecasts, the observed daily precipitation accumulations, a posterior predictive draw, and posterior predictive mean and standard deviation for a single day are shown in Figure \ref{fig:precip_ppred}. Observed and predicted zero accumulations are shown as gray. The figure shows good correspondence of the general spatial surface and smoothness characteristics in the observed and predicted precipitation amounts. 
Under the assumption that the library of observed historical climate states is sufficiently rich to match future states, our model can be used to make forecasts of future daily precipitation accumulations under various climate forcing scenarios. To do this, the 500 hPa geopotential height and surface temperature fields of the LENS RCP8.5 run are projected onto the first 10 principal components of their respective MERRA reanalysis fields. The distances between lagged trajectories of LENS EOF loadings and historical MERRA EOF loadings are calculated for all future days (e.g. for some future day $t'$, the distance vector $\mathbf{d}_{t'}$ consists of Euclidean distances between covariates on day $t'$ and all historical days $t = 1,\ldots, T$). LENS RCP8.5 forecasts of geopotential height and surface temperature are coupled with the fitted analogue model to make out-of-sample posterior predictive draws of precipitation on a $10 \mathrm{km}^2$ grid over three Susquehanna drainage basins from 2006-2100. Since the proposed model is defined on a continuous domain, predictions can be made on an arbitrarily fine grid. Summaries of the 3-month (April-May-June) maximum daily total precipitation volumes over basins are shown in Figure \ref{fig:rcp85_huc6_forecast}. The figure shows high uncertainty in the estimates, with 95\% credible bands capturing slight increasing trends over time for all three basins. To quantify the flood-risk, the forecasts from this model could be input into a hydrological flow model that accounts for topography and land use among other factors.

\section{Discussion} 
\label{sec:discussion}
While analogue methods have a long history in meteorology, there have been relatively few attempts at using them in a fully probabilistic framework for precipitation modeling. The analogue prior is a very general approach to modeling temporal dependence that leverages climate model forecasts of atmospheric variables that are concomitant with precipitation. Since this model is developed in a Bayesian framework, it is possible to account for uncertainty in the predictive skill of concomitant atmospheric variables that are used to select analogues. Moreover, the Student-t process mixture is a flexible model that can accommodate a wide variety of spatial dependence types in both the bulk and tail of the distribution. This model could be extended further by considering non-linear dimension reduction techniques for identifying analogues such as Laplacian eigenmaps \citep{Belkin02}, self organizing maps \citep{Kohonen12}, and diffusion maps \citep{Coifman06}. Similar to other analogue methods, the ability of the proposed model to produce accurate forecasts is predicated on the condition that there exist close historical analogues to future conditions \citetext{see \citealp{Vandendool94} for a discussion}. As part of our ongoing and future work, we plan to apply our precipitation forecasts to a hydrological water-flow model to more directly interrogate changes in flood-risk over the Susquehanna drainage basins. 

\appendix
\section{MCMC Details} 
\label{sec:mcmc_details}
Metropolis-Hastings MCMC algorithms were implemented for making posterior draws of the parameters in both the occurrence and intensity models using the R programing language (\href{http://ww.r-project.org}{http://ww.r-project.org}).

\subsection{Occurrence Model Metropolis-Hastings Algorithm} 
\label{sub:occurrence_model_metropolis_hastings_algorithm}
The parameters $\rho$, $\nu$, and $\theta$ were updated using variable-at-a-time random walks. Truncated normal Gibbs updates are available for the latent Gaussian process at observation locations $\mathbf{Z}_t = (Z_t(\bs_1), \ldots, Z_t(\bs_n))$. Denoting occurrence indicators $\mathbf{O}_t = (O_t(\bs_1), \ldots, O_t(\bs_n))$ and mean function $\boldsymbol{\mu}_t = (\mu_t(\bs_1), \ldots, \mu_t(\bs_n))$, the Gibbs updates for the latent Gaussian process are
\[
\mathbf{Z}_t|\mathbf{O}_t, \boldsymbol{\mu}_t, \rho, \nu \sim \mathrm{TN}_n(\boldsymbol{\mu}_t, \Sigma_{\nu, \rho}, \mathbf{l}_t, \mathbf{u}_t)
\]
where $\Sigma_{\nu, \rho}$ is the spatial covariance matrix for the $n$ observation locations, lower bounds $\mathbf{l}_t = (l(\bs_1), \ldots, l(\bs_n))$ have elements
\[
l(\bs_i) = \begin{cases}
	0, &\text{ if } O_t(\bs_i) = 1\\
	-\infty &\text{ if } O_t(\bs_i) = 0	
\end{cases}
\]
and upper bounds $\mathbf{u}_t = (u(\bs_1), \ldots, u(\bs_n))$ have elements
\[
u(\bs_i) = \begin{cases}
	\infty, &\text{ if } O_t(\bs_i) = 1\\
	0 &\text{ if } O_t(\bs_i) = 0
\end{cases}\qquad. 
\]
Gibbs updates are also available for both the location parameters $\gamma_t^{(O)}, t = 1, \ldots, T$ and $\boldsymbol{\beta}^{(O)}_t$. For the location parameter, the Gibbs update is
\[\resizebox{\linewidth}{!}{$
\gamma_t^{(O)}|\nu, \rho, \theta, \boldsymbol{\gamma}_{-t}^{(O)}, \sigma^2_\gamma, \boldsymbol{\beta}^{(O)}_t, \mathbf{Z}_t \sim \mathrm{N}\biggl\{\biggl(\mathbf{1}'\Sigma_{\nu, \rho}^{-1}[\mathbf{Z}_t - \boldsymbol{\psi}'\boldsymbol{\beta}^{(O)}_t] + \frac{\mathbf{w}_t'\boldsymbol{\gamma}_{-t}^{(O)}}{\sigma^2_{\gamma^{(O)}}}\biggr)\biggl(\mathbf{1}'\Sigma_{\nu, \rho}^{-1}\mathbf{1} + \frac{1}{\sigma^2_{\gamma^{(O)}}}\biggr)^{-1},\biggl(\mathbf{1}'\Sigma_{\nu, \rho}^{-1}\mathbf{1} + \frac{1}{\sigma^2_{\gamma^{(O)}}}\biggr)^{-1}\biggr\}.$}
\]
For the basis coefficients $\boldsymbol{\beta}^{(O)}_t$, the Gibbs update is
\[
\boldsymbol{\beta}^{(O)}_t|\mathbf{Z}_t, \gamma_t^{(O)}, \nu, \rho, \sigma^2_{\beta^{(O)}} \sim N_p\biggl\{\biggl(\boldsymbol{\psi}'\Sigma_{\nu, \rho}^{-1}\boldsymbol{\psi} + \frac{1}{\sigma_{\beta^{(O)}}^2}I_p\biggr)^{-1}\biggl(\boldsymbol{\psi}'\Sigma_{\nu, \rho}^{-1}[\mathbf{Z}_t - \gamma_t^{(O)}\mathbf{1}]\biggr), \biggl(\boldsymbol{\psi}'\Sigma_{\nu, \rho}^{-1}\boldsymbol{\psi} + \frac{1}{\sigma_{\beta^{(O)}}^2}I_p\biggr)^{-1}\biggr\}.
\]

Inverse Gamma gibbs updates are used for $\sigma^2_{\gamma^{(O)}}$ and $\sigma^2_{\beta^{(O)}}$. Using inverse Gamma parameterized with shape $a$ and scale $b$, having density $f(x; a, b) = \frac{b^a}{\Gamma(a)} x^{-(a + 1)} \exp\{- \frac{b}{x}\}, \, x > 0$. With prior $\sigma_{\gamma^{(O)}}^2 \sim \mathrm{IG}(a_\gamma,b_\gamma)$, the Gibbs update for $\sigma_{\gamma^{(O)}}^2$, is 
\[
\sigma_{\gamma^{(O)}}^2|\boldsymbol{\gamma}^{(O)}_t \sim \mathrm{IG}\biggl\{a_\gamma + n/2, b_\gamma + \frac{1}{2}\sum_{t =1}^T(\gamma_t - \mathbf{w}_t'\boldsymbol{\gamma_{-t}}^{(O)})^2\biggr\}, 
\]
and the Gibbs update for $\sigma^2_{\beta^{(O)}}$, assuming prior $\sigma_{\beta^{(O)}}^2 \sim \mathrm{IG}(a_\beta,b_\beta)$ is 
\[
\sigma_{\beta^{(O)}}^2|\boldsymbol{\beta}^{(O)}_{1:T} \sim \mathrm{IG}\biggl\{a_\beta + np/2, b_\beta + \frac{1}{2}\sum_{t = 1}^{T}\boldsymbol{\beta}^{(O)'}_t\boldsymbol{\beta}^{(O)}_t\biggr\}, 
\]
\subsection{Intensity Model Metropolis-Hastings Algorithm} 
\label{sub:intensity_model_metropolis_hastings_algorithm}
The parameters $\rho, \nu, a_k, b_k, \boldsymbol{\alpha_k}$ for $k = 1, \ldots, K$, and $\theta$ were updated using variable-at-a-time random walks. The cluster labels $\xi_t$ were also updated variable-at-a-time, but with discrete uniform independence proposals, each on 1:$K$. The location offset and basis functions Gibbs updates are similar to those in the occurrence model but are also dependent on the mixture labels.
\[\resizebox{\linewidth}{!}{$
\gamma_t^{(I)}|\nu_k, \rho_k, \theta, \boldsymbol{\gamma}_{-t}^{(I)}, \xi_t = k, \sigma^2_\gamma, \boldsymbol{\beta}^{(I)}_t, \mathbf{Z}_t \sim \mathrm{N}\biggl\{\biggl(\mathbf{1}'\Sigma_{k}^{-1}[\mathbf{Y}_t - \boldsymbol{\psi}'\boldsymbol{\beta}^{(I)}_t] + \frac{\mathbf{w}_t'\boldsymbol{\gamma}_{-t}^{(I)}}{\sigma^2_{\gamma^{(I)}}}\biggr)\biggl(\mathbf{1}'\Sigma_{k}^{-1}\mathbf{1} + \frac{1}{\sigma^2_{\gamma^{(I)}}}\biggr)^{-1},\biggl(\mathbf{1}'\Sigma_{k}^{-1}\mathbf{1} + \frac{1}{\sigma^2_{\gamma^{(I)}}}\biggr)^{-1}\biggr\}.$}
\]
where the covariance matrix $\Sigma_{k}$ is calculated using dependence parameters $\rho_k$ and $\nu_k$ corresponding to mixture class $k$.

Similarly, the basis coefficients $\boldsymbol{\beta}^{(I)}_t$ have Gibbs update
\[\resizebox{\linewidth}{!}{$
\boldsymbol{\beta}^{(I)}_t|\mathbf{Y}_t, \gamma_t^{(I)},\nu_k, \rho_k, \xi_t = k, \sigma^2_{\beta^{(I)}} \sim N_p\biggl\{\biggl(\boldsymbol{\psi}'\Sigma_{k}^{-1}\boldsymbol{\psi} + \frac{1}{\sigma_{\beta^{(I)}}^2}I_p\biggr)^{-1}\biggl(\boldsymbol{\psi}'\Sigma_{k}^{-1}[\mathbf{Y}_t - \gamma_t^{(I)}\mathbf{1}]\biggr), \biggl(\boldsymbol{\psi}'\Sigma_{k}^{-1}\boldsymbol{\psi} + \frac{1}{\sigma_{\beta^{(I)}}^2}I_p\biggr)^{-1}\biggr\}$.}
\]
The Gibbs updates for the prior variances $\sigma^2_{\gamma^{(I)}}$ and $\sigma^2_{\beta^{(I)}}$ are completely analogous to those in the occurrence model.


\bibliographystyle{spmpsci} 
\bibliography{biblio}

\end{document}